\documentclass[pra,twocolumn,reprint,showpacs,groupedaddress,superscriptaddress,nofootinbib,floatfix,preprintnumbers,longbibliography,aps]{revtex4-2}
\usepackage{xcolor}
\usepackage{graphicx}
\usepackage{dcolumn}
\usepackage{bm}

\begin{document}

\preprint{APS/123-QED}

\title{A spatiotemporal couplings perspective on harmonic vortices generation}

\author{C. Granados}
\email{camilo.granados@eitech.edu.cn}
\affiliation{Eastern Institute of Technology, Ningbo 315200, China}

\author{B. Kumar Das}
\affiliation{Department of Physics, Guangdong Technion - Israel Institute of Technology, 241 Daxue Road, Shantou, Guangdong, China, 515063}
\affiliation{Technion - Israel Institute of Technology, Haifa, 32000, Israel}
\affiliation{Guangdong Provincial Key Laboratory of Materials and Technologies for Energy Conversion, Guangdong Technion - Israel Institute of Technology, 241 Daxue Road, Shantou, Guangdong, China, 515063}

\author{M. F. Ciappina}
\email{marcelo.ciappina@gtiit.edu.cn}
\affiliation{Department of Physics, Guangdong Technion - Israel Institute of Technology, 241 Daxue Road, Shantou, Guangdong, China, 515063}
\affiliation{Technion - Israel Institute of Technology, Haifa, 32000, Israel}
\affiliation{Guangdong Provincial Key Laboratory of Materials and Technologies for Energy Conversion, Guangdong Technion - Israel Institute of Technology, 241 Daxue Road, Shantou, Guangdong, China, 515063}

\author{W. Gao}
\email{wgao@eitech.edu.cn}
\affiliation{Eastern Institute of Technology, Ningbo 315200, China}

\date{\today}

\begin{abstract}
The interaction of light with matter serves as a fundamental tool for probing material properties across a wide range of energy regimes. Recent breakthroughs in tailoring the topology of coherent electromagnetic fields have opened new avenues for exploring how matter uniquely responds to the topological characteristics of light. In this work, we conduct a comprehensive investigation of high-order harmonic generation (HHG) driven by spatiotemporal optical vortex (STOV) beams. We demonstrate how distinct STOV configurations imprint their signature on the HHG process and show that the intensity distribution of harmonic fields can be precisely controlled by tuning the beam parameters. Furthermore, by bridging microscopic calculations with far-field observations, we establish the consistency of our findings and offer fresh insights into this emerging nonlinear spatiotemporal regime.
\end{abstract}

\maketitle


\section{Introduction}
Coherent light has become an indispensable tool for probing and advancing our understanding of previously inaccessible non-perturbative phenomena \cite{AttoPhys}. These advancements now allow us to precisely manipulate individual particles, observe phenomena beyond the diffraction limit, and generate light in the extreme ultraviolet (XUV) regime \cite{AttoMicroscopy, AttoSci, QPheAtto, FemtoControl, VortexXUV}, among other breakthroughs. With the rapid progress in these applications and the increasing demand for more sophisticated tools to push scientific boundaries, it has become routine to control various properties of light, including its spin (SAM) and orbital angular momentum (OAM) \cite{VorBeam}. Light beams carrying OAM in the spatial domain are commonly referred to as vortex beams due to their geometric resemblance to vortices in classical and quantum fluids. These beams are characterized by phase singularities and a null field amplitude along a central axis (known as the vortex core), as well as their quantized OAM of $l\hbar$ per photon \cite{VorBeam}. In recent years, significant effort has been devoted to merging non-perturbative processes, such as high-order harmonic generation (HHG), with vortex beams \cite{BPOV, VortexXUV, StrongVort, TunableVort, ControlOAM}. HHG arises from the interaction of ultrashort light pulses with matter. Its underlying physical mechanism is well understood and explained by the three-step model, which describes HHG process as: (i) tunnel ionization, (ii) electron propagation, and (iii) recombination with the parent ion \cite{SFA}. 
HHG is the workhorse for the attosecond pulses generation, enabling the study of electron dynamics on their natural timescales \cite{Agostini1, Ros1, AttoMicroscopy, Keitel}. When HHG is driven by topological laser fields, it becomes possible to produce vortex beams in the XUV regime \cite{VortexXUV}, paving the way for the generation of twisted attosecond pulses—light pulses of attosecond duration that inherently carry OAM. Moreover, structured light holds significant potential for applications such as detecting chiral molecules \cite{ChiralMol, NonLinHeli}.

Nonetheless, unlike the well-established interest in light fields carrying OAM along the propagation direction \cite{PolMesurements, NonLinHeli, Tweezers, IntenseVHHG, AttoVortex}, light fields with transverse OAM (in the spatiotemporal domain) have been less explored in both the perturbative and non-perturbative regimes. The spatiotemporal optical vortex (STOV) beam \cite{STOV_NanoGratting,STOV_NatComm,STOV_NatPhot, STOV_PRX, MPorrasStov,weicaoAP} introduces an additional degree of freedom with vast potential for applications in light-matter interactions. For instance, chiral symmetry breaking in toroidal vortices of light \cite{ChiralToroidal} represents a significant advancement in this area, since it demonstrates the production of STOV with a chiral symmetry. Recently, STOV beams were used to drive harmonic generation using an 800~nm field focused in xenon gas \cite{STOV_Exp}. From a theoretical perspective, recent studies have demonstrated the potential to generate higher-order harmonics using STOV beams \cite{FreePropStov, PKU, STOVatto}. In addition, by introducing the quantum orbit version of the strong-field approximation (SFA), it was shown that the microscopic response of atoms and molecules to the topological light field could be investigated by directly introducing the beam structure into the SFA. This approach opens exciting new avenues for studying not only atomic and molecular systems but also solid state matter~\cite{solidPSTV}. However, the QO approach have, in our view, two weak points: (1) it is not possible to determine its validity since focusing a beam to the atomic size is not feasible, and (2) the interpretation of the harmonic generation process lacks a clear interpretation based on the characteristics of the driving field. 

In addition, several challenges arise when using vortex beams for HHG - a non-linear and non-perturbative process, whether in the spatial or spatiotemporal domains. For example, the core size of vortex beams expands with an increase in the topological charges (TCs). To address these issues, two types of beams with divergence independent of the TC have been proposed: (1) the non-diffractive Bessel STOV beam \cite{NSBessel}, where the beam size is maintained due to propagation effects, and (2) the perfect optical vortex (POV) beam \cite{BPOV}, whose divergence is entirely independent of the TC. The POV beam has been theoretically employed to demonstrate HHG in gases \cite{BPOV}, and its counterpart in the spatiotemporal domain has also been proposed \cite{PSTOV}. However, no studies have yet investigated the HHG process driven by the perfect spatiotemporal optical vortex (PSTOV) beams. Similarly, Bessel STOV beams have not been explored for their potential in HHG processes either. Additionally, Bessel STOV beams present a distinct framework for studying the impact of various field configurations on HHG. Their field distribution is characterized by the presence of  multiple concentric rings surrounding the main core, setting them apart from STOV or PSTOV beams. Investigating these complementary beam types is crucial for understanding the influence of different spatiotemporal beam parameters on the HHG process. 

In this work, we study HHG driven by STOV beams with a Gaussian envelope (GSTOV) and PSTOV beams. Using the GSTOV beam as a reference, we highlight the main differences between the two beams and analyze how their topological properties affect the HHG phenomena. By studying the HHG process in the near- and far-field regimes, we demonstrate: (1) the harmonic vortices chirp can be linked to the fundamental field's Fourier transform (FT), (2) Different STOV beams carry different spatiotemporal couplings, (3) the core size of the harmonics generated by the fundamental PSTOV beam show similar properties than the perfect spatial vortex beam, and (4) the lobed structure exhibited by the driving and harmonic fields, at least for the GSTOV case, is an intrinsic property of the conjugate character between the STOV beams and the Hermite polynomials. We will start by introducing different spatiotemporal beams and its properties in different Fourier domains. 

\section{Spatiotemporal vortex beams}
In this section, we will revisit the GSTOV and PSTOV beams, along with their Fourier transform to the spatiospectral domain, emphasizing the key characteristics and differences between these spatiotemporal vortex beams. All findings will be mainly described in the spatiospectral domain since it is customary to present HHG results in terms of the harmonic order. Notice that we define the chirp function to describe the central frequency of the individual lobes in the spatiospectral space.

\subsection{Gaussian-STOV}

\begin{figure}[h!]
\centering 
\includegraphics[width=1\linewidth]{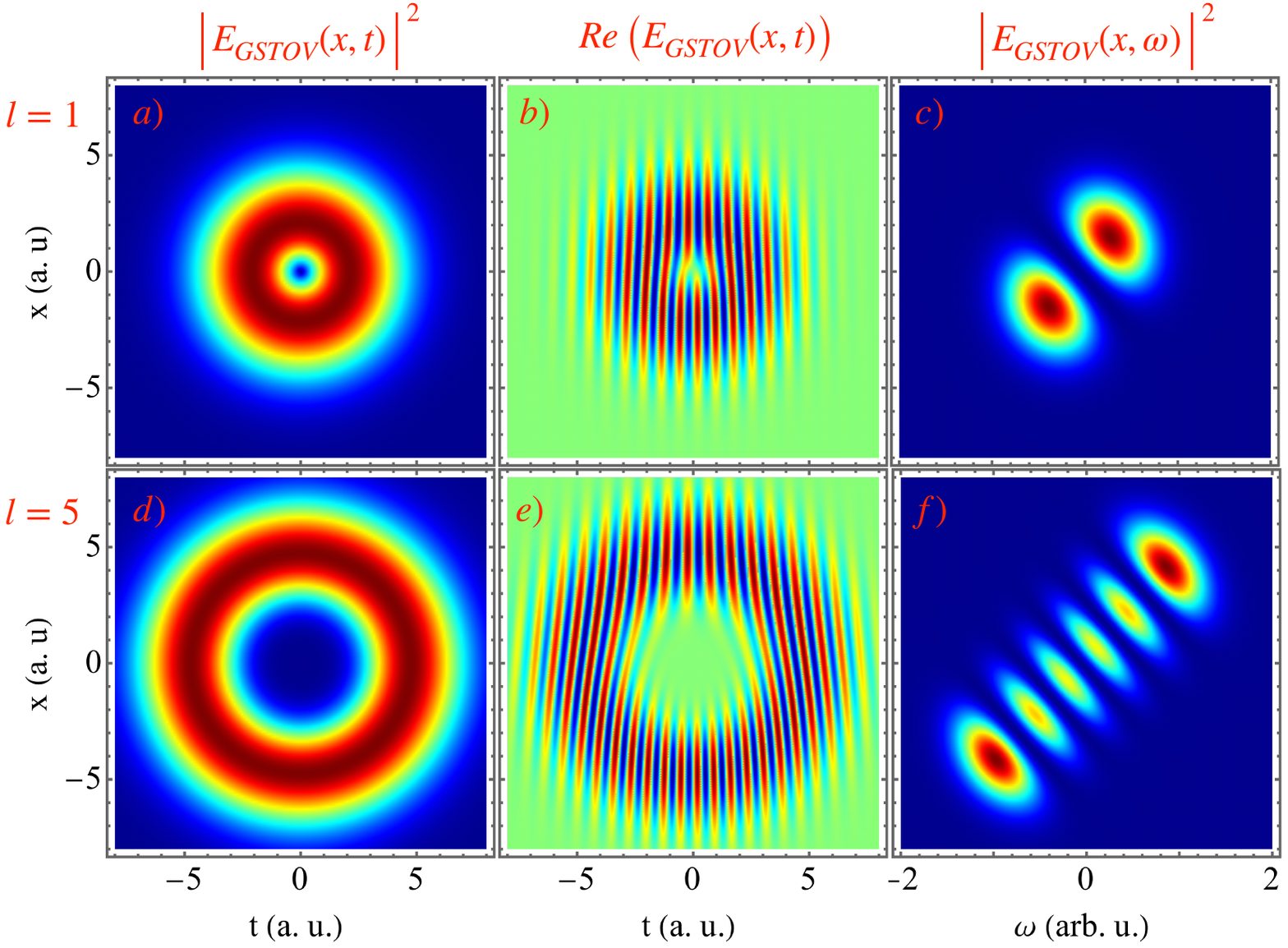}
\caption{Gaussian STOV beam features. (a) Intensity distribution, (b) real part and (c) FT in the spatiospectral domain, for a TC of $l$=1. (d)-(f) the same as (a)-(c), but with a TC of $l=5$.}
\label{Fig1}
\end{figure}

The GSTOV beam, corresponds to a Laguerre-Gauss vortex beam for a radial index $p=0$. Mathematically we can write the GSTOV beam as follows: 

\begin{eqnarray}
	E_{\mathrm{GSTOV}}(x,t) &=& A(x,t) \exp \left(-i \Phi \right),  \label{stbeam}
\end{eqnarray}
where $A(x,t) = E_0(\tau^2/w_\tau^2+x^2/w_x^2)^{|l|/2}\exp{(-x^2/w_x^2-\tau^2/w_\tau^2)}$ and $\Phi=  - l \phi - \omega_0 t - k_0 z $. The term $\phi = \arctan(\tau w_x / x w_\tau)$ denotes the spatiotemporal helical phase. For simplicity, we will investigate the STOV beam at the focus position, which allows us to neglect the contributions stemming from the Gouy phase and the phase-front curvature. The temporal coordinate $\tau$ is defined in terms of the moving time coordinate as $\tau = c (t - \delta t) - z$. The center of the pulse, $\delta t$, is fixed. The spatial and temporal pulse widths are represented by $w_x$ and $w_\tau$, respectively, and $l$ denotes the topological charge (TC). Unless specified otherwise, $w_\tau$ and  $w_x$ are set to $411$~$\mathrm{a.u.}$ and $3$~$\mathrm{a.u.}$, respectively. The driven laser wavelength and period are $\lambda = T_p c$ and $T_p = 2\pi / \omega_0$. Finally, the fundamental frequency is $\omega_0 = 0.057~\mathrm{a.u.}$. Figure~\ref{Fig1} depicts the main features of a GSTOV beam. In Fig.~\ref{Fig1}(a) we plot the intensity distribution, in Fig.~\ref{Fig1}(b) the real part and in Fig.~\ref{Fig1}(c) the FT in the spatiospectral domain. Here, the TC is $l=1$. In  Figs.~\ref{Fig1}(d)-(f), we show the same as in Figs.~\ref{Fig1}(a)-(c) but for a TC $l=5$. In all the panels, and for simplicity, we have used $y=z=0$. 

\subsection{Perfect-STOV}

\begin{figure}[h!]
\centering 
\includegraphics[width=1\linewidth]{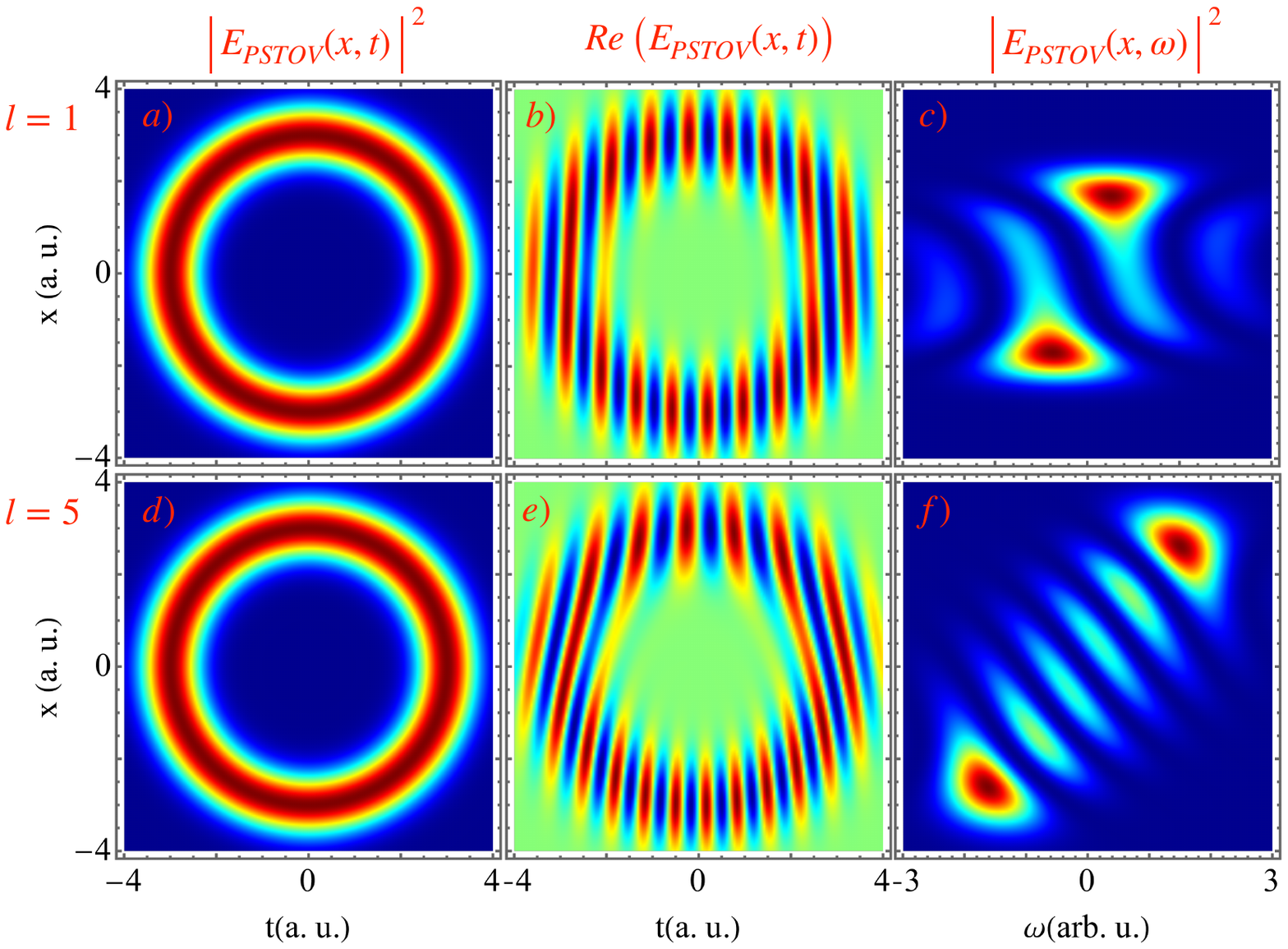}
\caption{Perfect STOV beam features. (a) Intensity distribution, (b) real part and (c) FT in the spatiospectral domain, for a TC of $l$=1. (d)-(f) the same as (a)-(c), but with a TC of $l=5$.}
\label{Fig2}
\end{figure}

\noindent In the spatial domain, a perfect optical vortex (POV) beam is characterized by a dark core and ring radius that remain independent of the TC. This property provides a significant advantage over other vortex beams, such as the Laguerre-Gauss or Bessel-Gauss beams, where the dark core size increases monotonically with the TC. Although HHG can produce harmonics with high TCs, the process is typically limited to driving fields with low TC values. This limitation arises due to a trade-off between the peak intensity of conventional vortex beams and the TC—higher TC values result in a significant decrease in peak intensity, making the HHG process unfeasible. Using a POV beam to drive the HHG process overcomes this limitation, enabling the generation of harmonics with very high TCs~\cite{BPOV}. The spatiotemporal counterpart of the POV beam, the PSTOV beam, retains the core properties of the spatial POV beam while also carrying an orbital angular momentum (OAM) perpendicular to the beam’s propagation direction, known as the transverse OAM. Mathematically, the PSTOV beam can be described using the modified Bessel function of the first kind \cite{PSTOV}. As in the spatial case, the asymptotic form of the modified Bessel function (under the condition $r_{0}\gg \delta$, where $r_0$ is the vortex ellipse radius and $\delta$ the width of the vortex's maximum intensity region) can be used to approximate the field distribution of the PSTOV beam~\cite{PSTOV}:
\begin{eqnarray}
    E_{\mathrm{PSTOV}}(x,t)&=& E_0 \exp \left(-i \Phi \right)\exp\left(- \frac{(r-r_0)^2}{\delta^2 } \right), \label{pstov}
\end{eqnarray}
where $r = \sqrt{x^2/w_x^2 + \tau^2/w_\tau^2}$ represents the spatiotemporal coordinate. The vortex ellipse radius is set to $r_0 = 1~\mathrm{a.u.}$, and the width of the vortex’s maximum intensity region is $\delta = 0.3~\mathrm{a.u}$. Additionally, the beam parameters are kept the same as for the case of the GSTOV beam. Figure~\ref{Fig2} shows the main PSTOV characteristics. In Fig.~\ref{Fig2}(a), we present the intensity distribution, in Fig.~\ref{Fig2}(b) the real part , and in Fig.~\ref{Fig2}(c) the FT in the spatiospectral domain and for a TC value $l=1$. In  Figs.~\ref{Fig2}(d)-(f), we show the same as in Figs.~\ref{Fig2}(a)-(c) but for a TC $l=5$. 

\subsection{Spatiotemporal couplings in the fundamental beam}

A very important characteristic defining the STOV beams is the non-separability of their spatial and temporal parts, i. e., $E_{\mathrm{STOV}}(x,t) \ne E(x)  E(t)$, given by the helical phase. The spatial and temporal parts of the field are intrinsically  entangled with each other that gives rise to the spatiotemporal couplings (STCs) \cite{ST_Control1,ST_Control2,ST_Control3}. The STCs manifest in different Fourier domains as lobed chirped intensity distributions \cite{trevino,RhodesSTC, HullierSTC}. The chirp of different spatiotemporal beams can be expressed as a function of $\omega$ and can be extracted from the FT to the spatiospectral domain.

\begin{figure}
\centering 
\includegraphics[width=1\linewidth]{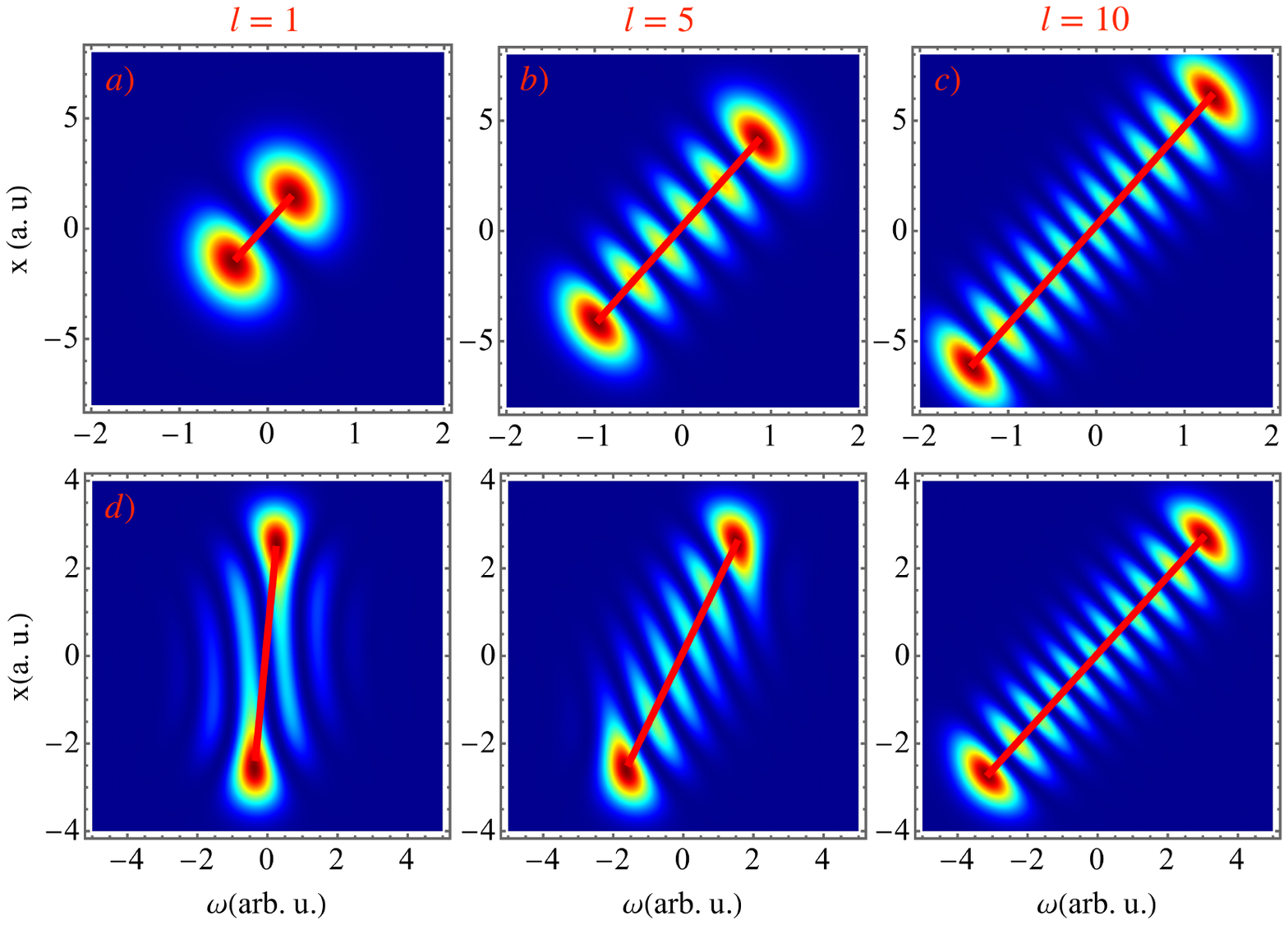}
\caption{Intensity distributions of the GSTOV and PSTOV  beams for different topological charges in the spatiospectral domain. In (a) to (c) the GSTOV beam for TC values $l=1,5$, and 10. In (d) to (f) the same as before but this time for the PSTOV. The intensity distribution specially for a TC value $l=1$, clearly differs from the GSTOV beam. In addition, it is clear that the larger the TC value, the closer the PSTOV to the GSTOV, as shown in panels (c) and (f). }
\label{Fig3}
\end{figure}

In Fig.~\ref{Fig3}, we show the intensity distribution of the GSTOV beam for TC values of (a) $l=1$, (b) $l=5$, and (c) $l=10$ in the spatiospectral domain. In panels (d) to (f), we use the same TC values but for the PSTOV beam. In all the panels, the straight line represents the chirp function extracted from the FT of the fundamental field mapping to the spatiospectral domain. For the case of the GSTOV beam, the FT is given by (for $l=1$) \cite{MPorrasStov}: 

\begin{eqnarray}
	\widetilde{E}_{\mathrm{GSTOV}}(x,\omega) &\propto&  -\frac{iE_0}{ \sqrt{2}} \exp\left( -\frac{x^2}{\omega_x^2} \right) \exp\left( -\frac{\omega_\tau^2}{x^2} (\omega-\omega_0)^2 \right)  \nonumber \\ 
	&\times& \Big( x - \frac{\omega_\tau \omega_x}{2c}( \omega- \omega_0)  \Big).
\label{FTGSTOV}
\end{eqnarray}

To extract the chirp function, we make Eq.~(\ref{FTGSTOV}) equal to zero and solve for $x$. To our knowledge the PSTOV beam's FT has no analytical solution, however, by using a similar expression to the one given by Eq.~(\ref{FTGSTOV}), we derive an expression for the PSTOV chirp function. The chirp function can be written as follows: 
\begin{eqnarray}
	x_{G}(\omega)&=& \frac{w_x  w_\tau}{2c} \omega  \nonumber \\
	x_{P}(\omega)&=& \frac{w_x  w_\tau}{1.05 l  c} \omega , \label{STC_Fundamental}
\end{eqnarray}
with $x_{G}(\omega)$ and $x_{P}(\omega)$ representing the chirp functions for the GSTOV and PSTOV beams, respectively. Interestingly, the PSTOV beam is more sensitive to the TC than the GSTOV beam, which is evident from Eq.~(\ref{STC_Fundamental}) with $x_{P}(\omega) \propto 1/l$. This explains why in the spatiospectral domain, the intensity distribution, Figs.~\ref{Fig3}(e) and (f), exhibits a smaller inclination for larger TC values. 

We can conclude that the STCs dictate the chirp function and intensity distributions of the fundamental field in the spatiospectral domain. Additionally, by solving numerically the FT, we observe clear differences between the PSTOV and GSTOV chirp functions and intensity distributions.

\section{Harmonic Vortices}

Fundamentally, the HHG process describes the interaction of an intense ultrashort laser field with a target medium (atoms, molecules and solids). Briefly, the process in atoms and molecules consists of three main steps: (1) tunnel ionization of an electron by the driving laser field, (2) acceleration of the ionized electron in the continuum by the laser field, and (3) recombination of the electron with the parent ion core and releasing the acquired energy as a high harmonic photon. To calculate the harmonic response of the target medium to an STOV beam, one can assume different types of laser fields that vary only in time, as the dipole approximation allows the spatial dependence to be neglected. This simplification arises from different spatial scales involved between the target and the driving laser field, i.e.~the spatial distribution of the vortex beam is much larger than the electron excursion distance in the continuum. Typically, the macroscopic description of the HHG process driven by spatial optical vortices is performed using the so-called thin-slab model (TSM)~\cite{VortexXUV, TSM}. Using this model, it becomes easier to understand how the vorticity of the fundamental field is transferred to the harmonic fields, as it provides a far-field description. Furthermore, the model allows the spatial distribution of the electromagnetic field and the temporal response of the target system to be treated independently. For spatiotemporal vortex beams, the temporal dynamics of the field can be incorporated into the strong-field approximation (SFA)~\cite{SFA} to compute the harmonic response of the target while still leveraging the dipole approximation. This approach is commonly referred to as the quantum orbit (QO) approach \cite{PKU,STOVatto,QO}. The QO version of the SFA (QO-SFA) has also been applied to study spatial toroidal vortices \cite{SFA_OV} to generate exotic structures. However, such studies remain experimentally unfeasible for a simple reason: the spatial transverse structure of the vortex beam will always be many times larger than the size of the target atom. In addition, recently \cite{GranadosPRL} we demonstrated that the QO approach leads to wrong STCs in the near-field. In this section, we will start presenting the harmonic vortices in the near- and far-fields followed by the far-field calculations. FFor both cases, we will discuss the impact of the STCs on the harmonic field. Additionally, we will restrict our analysis to the harmonic orders $17^{\text{th}}$ to $21^{\text{st}}$, since for these harmonics the scaling laws are similar \cite{VortexXUV}.  

\subsection{Near-Field harmonic vortices}

The TSM \cite{VortexXUV, TSM} allows for the calculation of the near-field amplitude and phase of a particular harmonic by first expressing the harmonic amplitude as the fundamental field amplitude raised to a constant power given by the power scaling law of the harmonic with the fundamental beam power, and the harmonic phase as the harmonic order times the fundamental field's phase. In general, for a driving STOV beam, amplitude of the $q^{\text{th}}$ order harmonic in the near-field is given by, $E_{\mathrm{STOV}}^{\mathrm{near}}(x,t)=|A(x,t)|^p \exp(-iql\phi-iq\omega_0 t-i\alpha I(x,t))$, with $\alpha I(x,t)$ being the intrinsic phase of the $q^{\text{th}}$ order harmonic. Here, $I(x,t)$ is the fundamental beam intensity and $\alpha$ is a quantum path-dependent strong-field parameter~\cite{Balcouphase}. In our theoretical calculation, we neglect the dipole phase contribution to the harmonic generation process. This is due to the fact that the dipole phase plays a minor role with the driving STOVs, which has also been demonstrated experimentally~\cite{STOV_Exp}. In Ref.~\cite{STOV_Exp}, it is shown that the role of the dipole phase is only to bend the harmonic stripes in the far-field spectrum. The intensity of the near-field harmonics in the spatiospectral domain can be calculated as follows: 
\begin{eqnarray}
    \left|\widetilde{E}^{\mathrm{near}}_{\mathrm{STOV}}(x,\omega)\right|^2 &\propto& \Bigg|\int dt \left(A(x,t)\right)^p \exp{\left( - i q l \phi \right)}\nonumber \\
    &\times& \exp{\left(-i q t(\omega+\omega_0) \right)}\Bigg|^2,
\end{eqnarray}
\begin{figure}[h!]
\centering 
\includegraphics[width=1\linewidth]{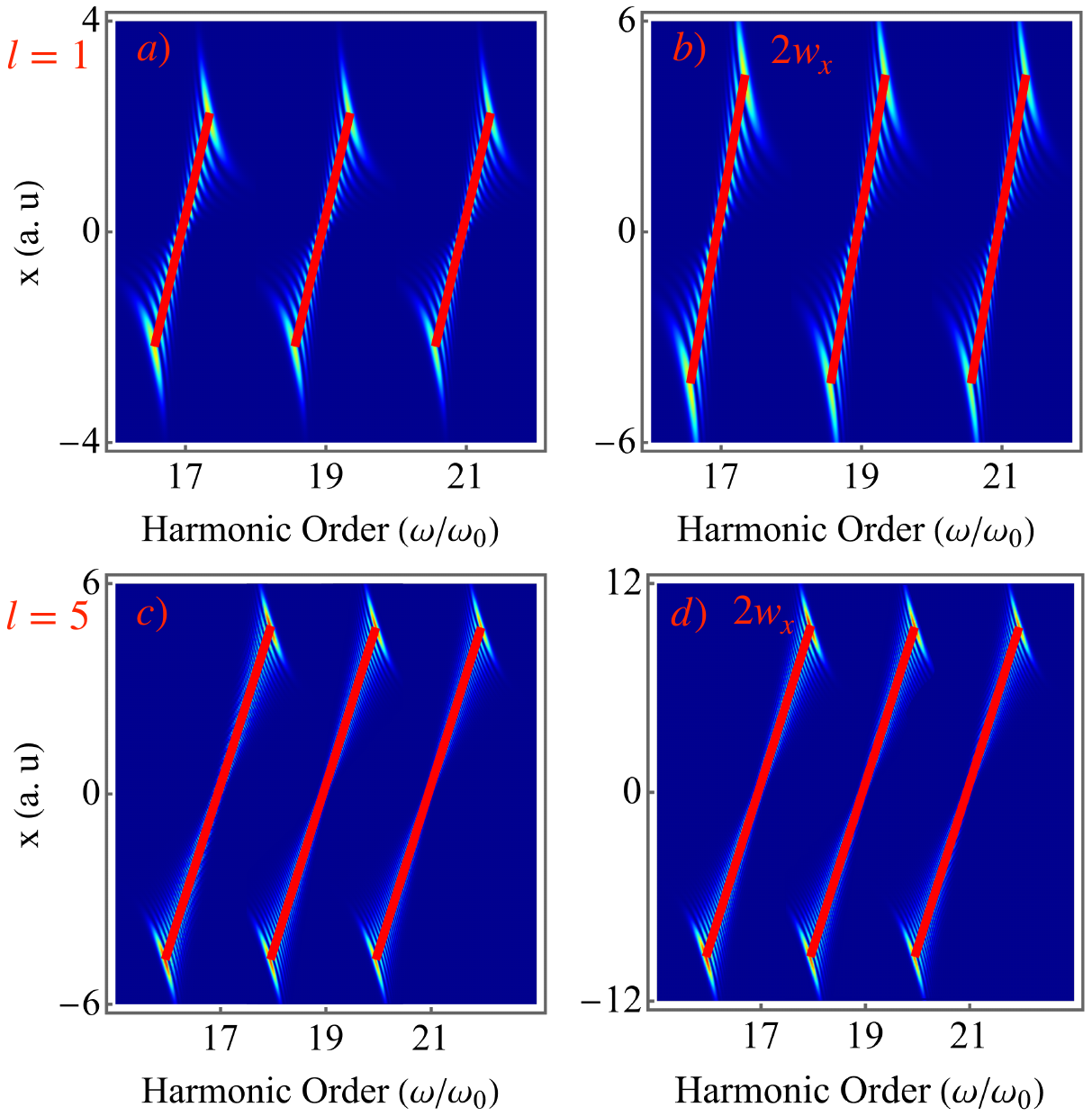}
\caption{Near-field harmonic vortices in the spatiospectral domain driven by a GSTOV beam . In (a) and (b), we show the GSTOV beam generated harmonic orders $17^{\text{th}}$ to $21^{\text{st}}$ with TC value $l=1$ and two different spatial widths ($w_x=3\mathrm{a.u.}$). In (c) and (d), the same harmonic vortices for a TC value $l=5$ and two different widths. }
\label{Fig4}
\end{figure}
where $p$ represents a scaling parameter extracted from the SFA simulations. In Fig.~\ref{Fig4}, we show the harmonic vortices with orders $17^{\text{th}}$ to $21^{\text{st}}$ for a driving GSTOV beam. In Figs.~\ref{Fig4}(a)-(b) we set a TC of $l=1$, and in Figs.~\ref{Fig4}(c)-(d) we set a TC of $l=5$. Note that the $x$-axis scale in Figs.~\ref{Fig4}(b) and (d) is larger. In all the panels, the red line represent the chirp function given by $x_{GNF}(\omega)= w_x w_\tau \omega/nc$.  The subindex $GNF$ indicates that the GSTOV beam is driving the harmonic generation process and that we are observing them in the near-field. In Figs.~\ref{Fig4}, (a) and (b), $n=1.6$, and in Figs.~\ref{Fig4}(c) and (d) we use $n=1.9$, which shows a small effect of the TC, since for the fundamental field in the $x-\omega$ domain, $n=2$ (see Eq~(\ref{STC_Fundamental})). Analyzing the different panels, it is clear that: (1) for harmonic vortices, generated by the GSTOV beam, the STCs are little sensitive to the TC effects, and (2) as expected, changing  the driving field width induces changes in the STCs. This is evident from panels (b) and (c) by noticing a larger $x$-axis scale.

\begin{figure}[h!]
\centering 
\includegraphics[width=1\linewidth]{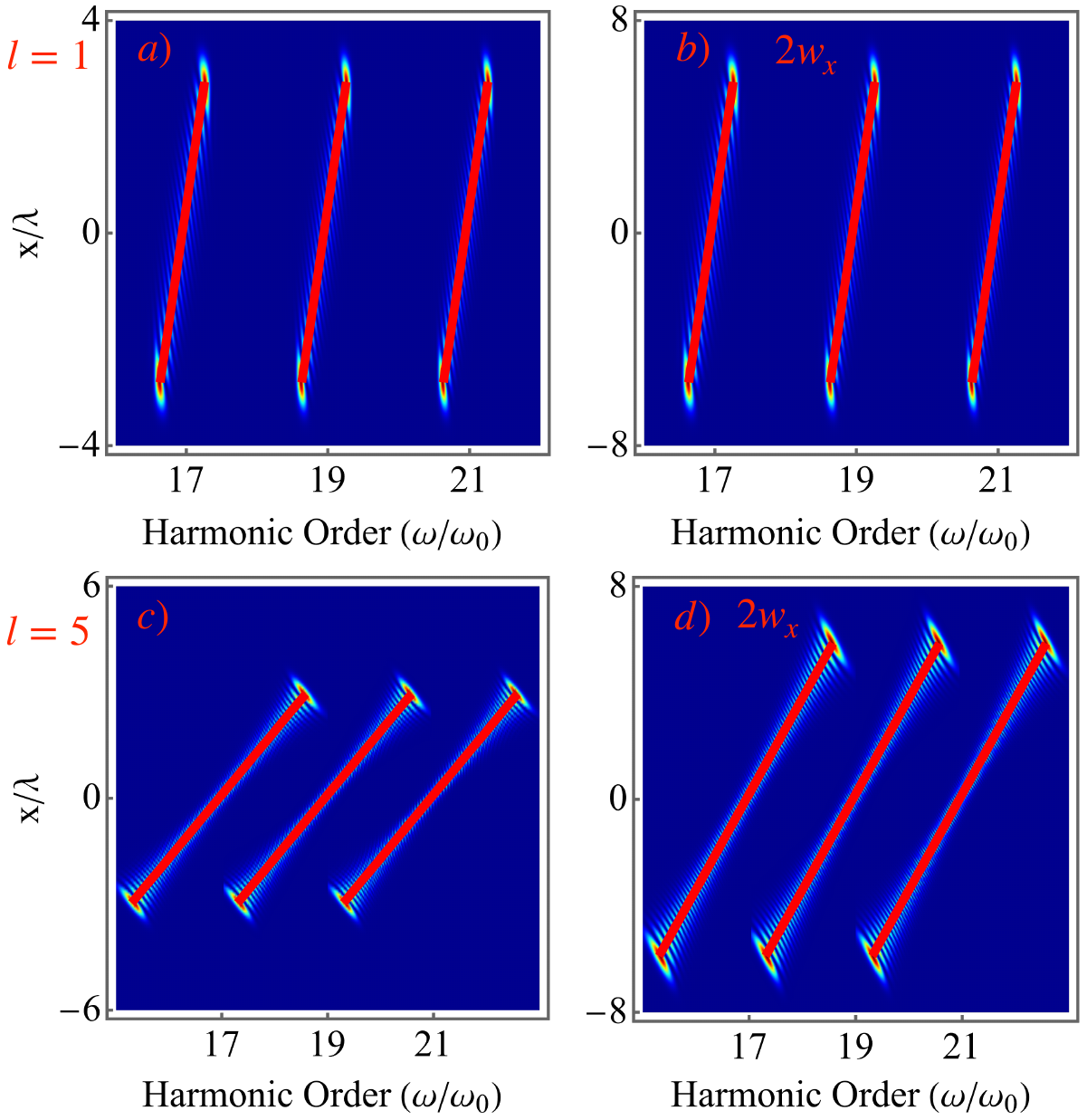}
\caption{Near-field harmonic vortices in the spatiospectral domain driven by a PSTOV beam. In (a) and (b), we show the PSTOV beam generated harmonic orders $17^{\text{th}}$ to $21^{\text{st}}$ with TC value $l=1$ and two different spatial widths  ($w_x=3\mathrm{a.u.}$). In (c) and (d), the same harmonic vortices for a TC value $l=5$ and two different widths. }
\label{Fig5}
\end{figure}

We present the numerical results for the PSTOV case in Fig.~\ref{Fig5}. In panels (a) and (b), we show the harmonic vortices in the spatiospectral domain for a TC value of $l=1$ and two different spatial widths. In (a), the spatial width is $3$~$\mathrm{a.u.}$, and in (b), $6$~$\mathrm{a.u.}$. For panels (c) and (d), we show the same as before but for a TC value of $l=5$. In all the panels, the red straight lines represent the chirp function $x_{PNF}(\omega)= w_x w_\tau \omega/nlc$, with $n=1.05$ for panels (a) to (b) and $n=1.0$ for panels (c) to (d). The subindex $PNF$ indicates that the PSTOV beam is driving the harmonic generation process and that we are observing them in the near-field. From different panels we can, once again, conclude that the characteristics of the fundamental field are imprinted into the  harmonic vortices and are fundamentally a manifestation of the STCs. Notice that, the harmonic field STCs are slightly different to the fundamental field, which can be explained by the scaling law used to calculate the near-field. Differently, for the FT of the fundamental field, the scaling parameter is 1. In addition, the near-field harmonic vortices chirp exhibit an inverse dependence on the TC, i.e., $x_{PNF}(\omega)\propto1/l$. Additionally, for both GSTOV and PSTOV, there is no dependency of the chirp function on the harmonic order for any of the scenarios explored in the near-field. Finally, the conservation of the TC in the harmonics is clear from the number of minima between the lobes in the intensity distribution. 

\subsection{Far-Field harmonic vortices}

\begin{figure}[h!]
\centering 
\includegraphics[width=1\linewidth]{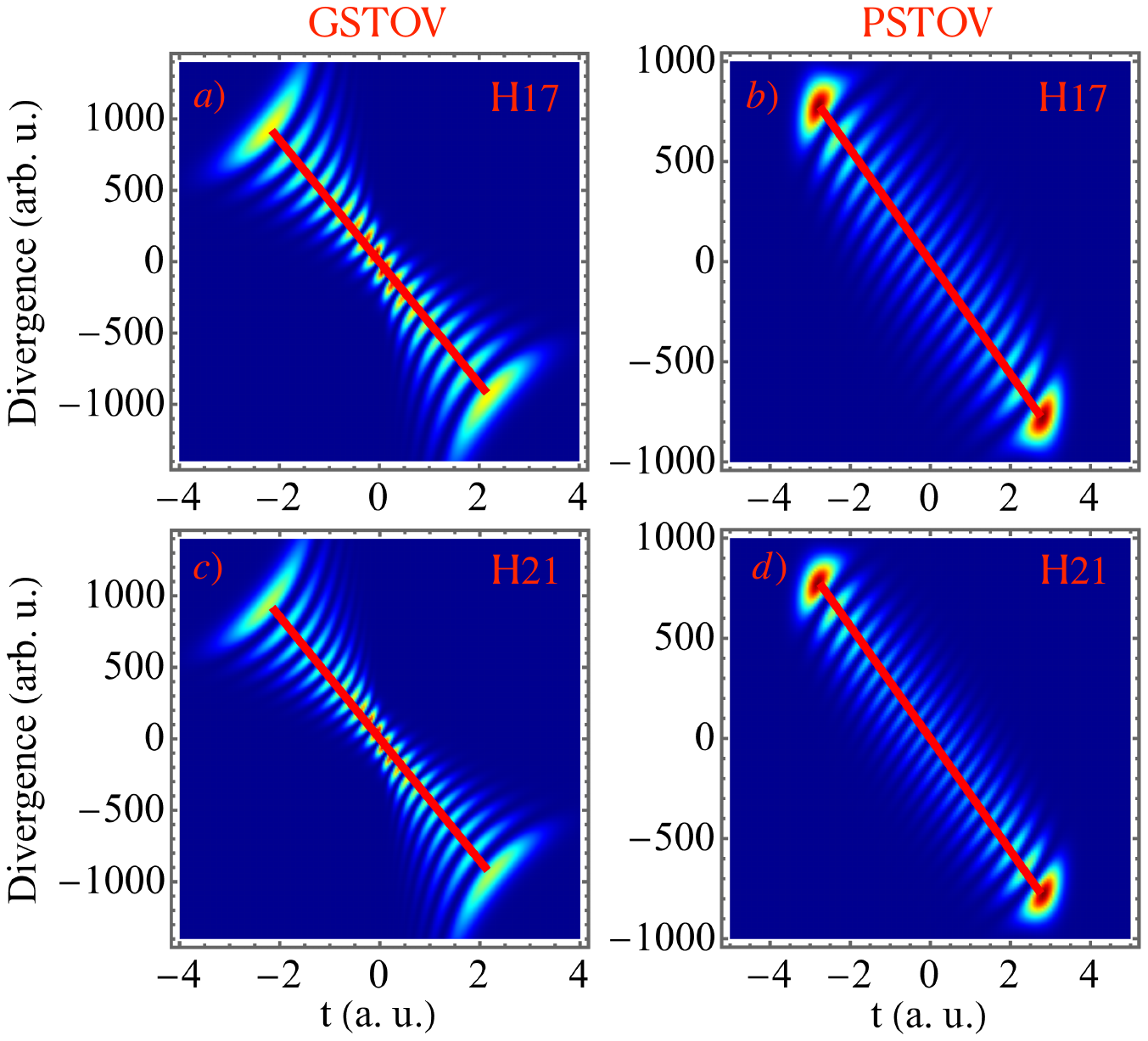}
\caption{Far-field harmonic vortices in the $\beta_x-t$ domain. In panels (a) and (b), we show the intensity distribution of the 17$^{\text{th}}$ harmonic order generated by the GSTOV and the PSTOV beams, respectively. In panels (c) and (d), we show the intensity distribution of the 21$^{\text{st}}$ harmonic order generated by the GSTOV and the PSTOV beams, respectively.}
\label{Fig6}
\end{figure}

For calculating the harmonic amplitude and phase in the far-field, the Fraunhofer integral combined with the near-field harmonics can be used. Notice that the integral should be solved for the two far-field coordinates, here represented by $\beta_x$ and $\beta_y$, where, $\beta=\sqrt{\beta_{x}^2+\beta_{y}^2}$ is the far-field divergence. However, in order to see the effect of the STC in the harmonic field intensity distribution and the chirp, we restrict the results to solve the integral only for $\beta_x$. The harmonic field intensity in the far-field is represented by the following integral: 
\begin{eqnarray}
   \Bigg| \widetilde{E}^{\mathrm{far}}_{\mathrm{STOV}}(\beta_x,t)\Bigg|^2 &\propto&\Bigg| \int dx E^{\mathrm{near}}_{\mathrm{STOV}}(x,t) \exp{\left( - i q l \phi \right)}\nonumber \\
    &\times& \exp{\left(-i \omega_q t \right)}\exp{(-ik_q \beta_x x)}\Bigg|^2,\label{ffintegral}
\end{eqnarray}
with $k_q$ representing the wavenumber of the $q^{\text{th}}$ order harmonic and $ \omega_q =  q\omega_0$. In Fig.~\ref{Fig6}, we show the 17$^{\text{th}}$ and 21$^{\text{st}}$ harmonic vortices in the $\beta_x-t$ (divergence-time) domain. The intensity distribution shows two important characteristics: (1) The conservation of the orbital angular momentum, as for the near-field case, is manifested by the number minima between the lobes. This is evident for both the GSTOV and PSTOV beams. (2) The STCs in the $\beta_x-t$ domain are given by: 
\begin{eqnarray}
\beta_x = - \frac{t n c}{w_x w_\tau k_0}. \label{FF_chirp}
\end{eqnarray}
In Eq.~(\ref{FF_chirp}), $n=1.6$ and $n=1.05 l$ for the GSTOV and PSTOV beams, respectively. Additionally, the constant value changes, as for the near-field, for large values of TC (not shown here). It is clear that the FT of the fundamental field can be used to explain the chirp in the harmonic vortices and the conservation of OAM explains the intensity distribution of different harmonic vortices. This is a significant result, as some interpretations found in the literature may be misleading~\cite{PKU}. 

\begin{figure}[h!]
\centering 
\includegraphics[width=1\linewidth]{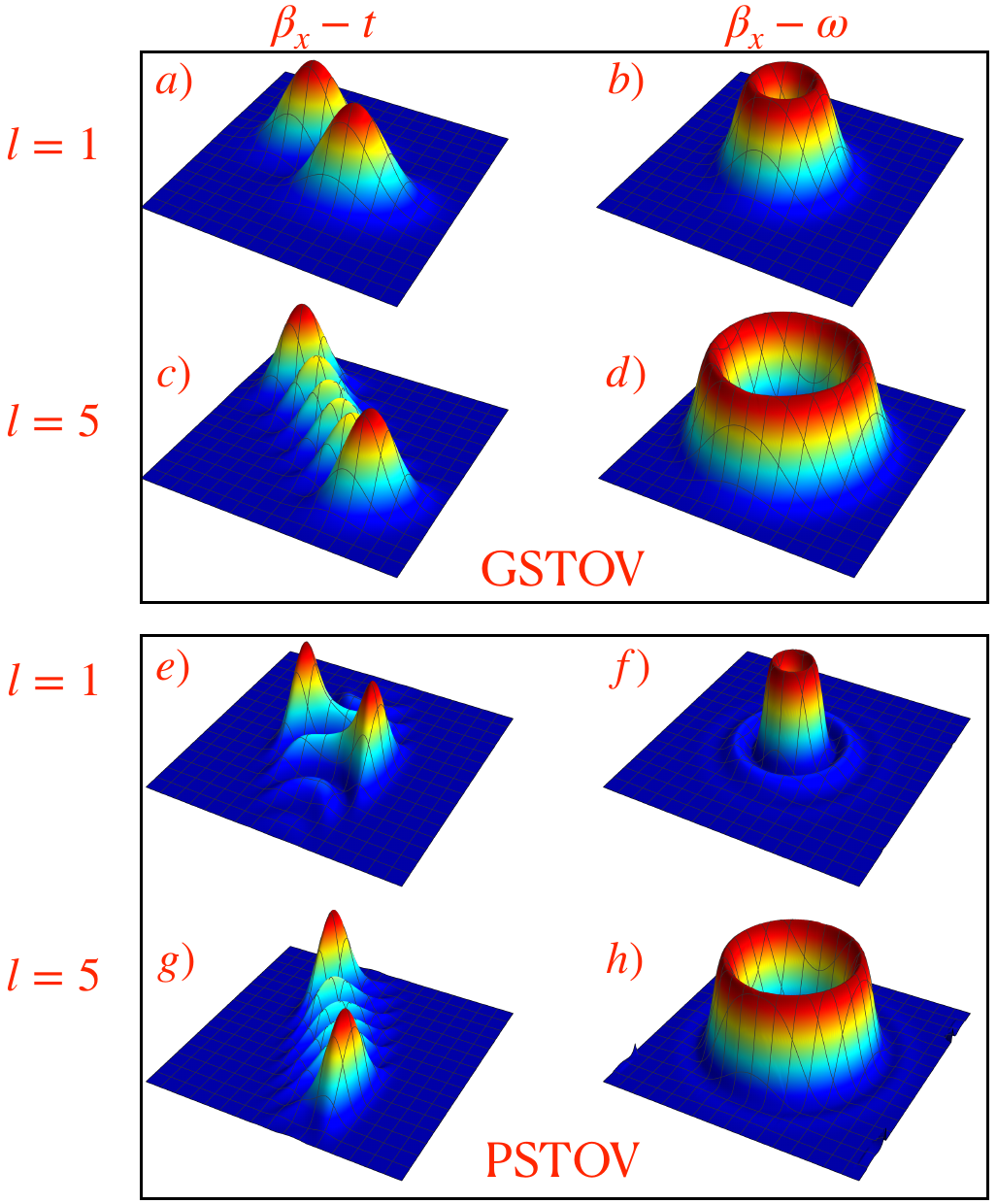}
\caption{Far-field representation of the fundamental GSTOV and PSTOV beams in the $\beta_x-t$ and $\beta_x-\omega$ domains. In all the plots, $\beta_x$ is the presented in the vertical axis, and $t$ and $\omega$ are presented in the horizontal axis. }
\label{Fig7}
\end{figure}

In addition to the far-field description of the HHG process in the $\beta_x-t$ domain, it is important to describe the process also in the frequency domain i.e., in the $\beta_x-\omega$ domain. We start here by showing the fundamental field. Firstly, we numerically solve the integral in the far-field and later perform a FT from the temporal domain, $t$, to the frequency domain, $\omega$. In Fig.~\ref{Fig7}(a)-(d), we show the GSTOV beam for TC values $l=1$ and 5. In Fig.~\ref{Fig7}(e)-(h), we show the PSTOV beam for the same TC values. 

From Fig.~\ref{Fig7}, we observe several fundamental characteristic features of the spatiotemporal vortices: (1) The GSTOV beam in the $\beta_x-\omega$ domain shows a larger radius of the maximum intensity for larger TC values, a characteristic feature that is typically found in the spatially structured light beams. It also clear that such characteristic feature is transferred to the spatiotemporal domain as well \cite{STOV_NatPhot}. (2) The PSTOV beam in the far-field shows an onset of the secondary rings around the main core, as clearly noticed in panels (f) and (h). This is expected since the perfect optical vortex and the Bessel-Gauss vortex beams are conjugate pairs \cite{BPOV}. 

\begin{figure}[h!]
\centering 
\includegraphics[width=1\linewidth]{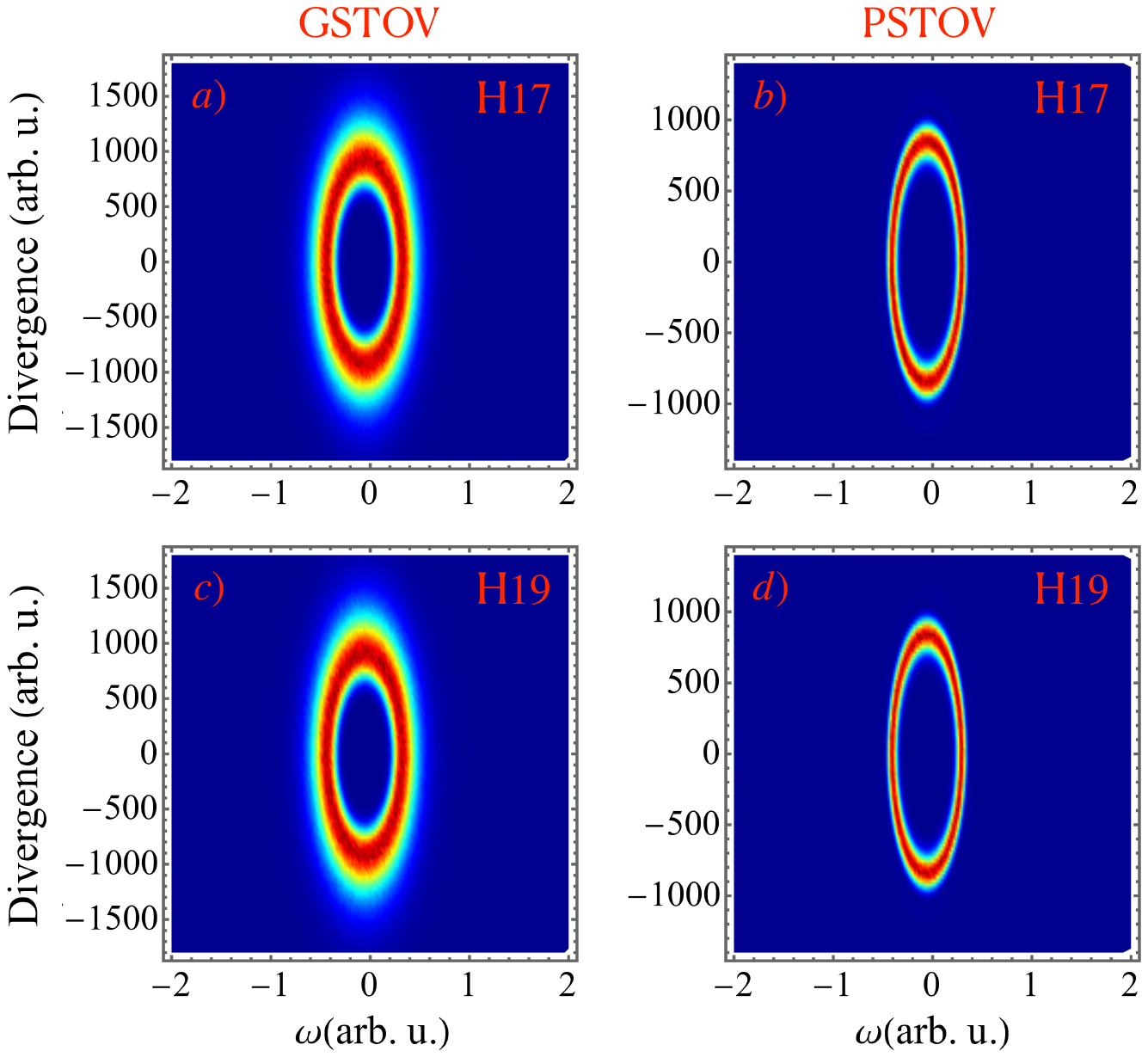}
\caption{Far-field harmonic vortices driven by the GSTOV and PSTOV beams. All the plots are displayed in arbitrary units. In panels (a) and (b), we show the intensity distribution of H17 in the $\beta_x-\omega$ domain. In panels (c) and (d), the same as before but for H19. }
\label{Fig8}
\end{figure}

In Fig.~\ref{Fig8}, we show the generation of 17$^{\text{th}}$ and 19$^{\text{th}}$ harmonic orders in the $\beta_x-\omega$ domain, when HHG is driven by the fundamental GSTOV (see (a) and (c)) and PSTOV (see (b) and (d)) beams. Figures in the $\beta_x-\omega$ domain reveal several important characteristics: (1) The ring thickness of harmonic vortices is larger when the harmonic generation is driven by the GSTOV beam rather than the PSTOV beam, showing that the differences exhibited by the fundamental fields (Figs.~\ref{Fig1} and~\ref{Fig2}) are translated to the harmonics. This has already been demonstrated for spatial perfect vortex beam \cite{BPOV}, (2) The harmonic intensity distribution for both beams exhibit nearly similar divergences in the far-field. Such an important characteristic has also been demonstrated when HHG is driven by a spatial vortex beams\cite{TSM, BPOV}, which is instrumental in generating twisted attosecond pulses. Our theoretical results also demonstrate the same characteristic for the spatiotemporal vortex beam driven HHG \cite{STOV_Exp}, making the PSTOV beam an important candidate to generate XUV beams with high TCs.

\section*{Conclusions}

We studied the harmonic generation process driven by spatiotemporal vortex beams to clarify several misleading interpretations presented in the literature. By analyzing two different fundamental fields, we demonstrate that spatiotemporal couplings determine both the intensity distribution and the characteristic chirp exhibited in the $x-\omega$ and $\beta_x-t$ Fourier domains. This was corroborated by showing that, for different values of the topological charges, the chirp of the beam can still be described using spatiotemporal couplings. Furthermore, we showed that characterizing the fundamental field allows for a proper description of both the intensity distribution and the chirp function of the generated harmonics. For both Gaussian and perfect spatiotemporal vortex beams, we derived mathematical expressions to describe the chirp function. These functions for the harmonic vortices are closely related and can be derived from the couplings exhibited by the fundamental field. One key difference between Gaussian and perfect beams was observed: the chirp function for the perfect spatiotemporal vortex beam depends on the topological charge of the fundamental beam. Additionally, it is clear that for both beam types, the chirp does not depend on the harmonic number. This is significant because it establishes the correct approach for calculating harmonic vortices in both the near- and far-fields. Finally, by analyzing the harmonics in the far-field, we demonstrated that characteristics of the driving fields are imprinted onto the harmonic vortices. Notably, we showed that the perfect spatiotemporal vortex beam is an ideal candidate for generating harmonics with short wavelengths and large topological charges, as the fundamental beam's radius of maximum intensity remains unchanged with varying topological charge.

In summary, we demonstrated that the two main characteristics exhibited by harmonics in the spatiospectral domain—the lobed intensity distribution and the chirp—can be explained by the Fourier transform of the fundamental beam rather than by multiple interference sources, as previously suggested in the literature.

\section*{Acknowledgments} The present work is supported by the National Key Research and Development Program of China (Grant No.~2023YFA1407100), Guangdong Province Science and Technology Major Project (Future functional materials under extreme conditions - 2021B0301030005) and the Guangdong Natural Science Foundation (General Program project No.~2023A1515010871).

\bibliography{apssamp}

\end{document}